\documentclass[conference]{IEEEtran}
\IEEEoverridecommandlockouts
\usepackage{cite}
\usepackage{amsmath,amssymb,amsfonts}
\usepackage{algorithmic}
\usepackage{graphicx}
\usepackage{textcomp}
\usepackage{caption}
\usepackage{subcaption}
\usepackage{comment}
\usepackage{xcolor}
\usepackage{multirow}
\usepackage{multicol}
\usepackage{makecell}
\usepackage[hyphens]{url}
\usepackage{hyperref}
\def\BibTeX{{\rm B\kern-.05em{\sc i\kern-.025em b}\kern-.08em
    T\kern-.1667em\lower.7ex\hbox{E}\kern-.125emX}}
\begin{document}

\title{COVIDFakeExplainer: An Explainable Machine Learning based Web Application for Detecting COVID-19 Fake News\\
\thanks{\copyright2023 IEEE. This manuscript has been accepted to publish in the 10th IEEE Asia-Pacific Conference on Computer Science and Data Engineering (IEEE CSDE 2023). Personal use of this material is permitted.  Permission from IEEE must be obtained for all other uses, in any current or future media, including reprinting/republishing this material for advertising or promotional purposes, creating new collective works, for resale or redistribution to servers or lists, or reuse of any copyrighted component of this work in other works.}
}

\author{\IEEEauthorblockN{Dylan Warman}
\IEEEauthorblockA{\textit{School of Computing, Mathematics and Engineering} \\
\textit{Charles Sturt University}\\
NSW, Australia \\
dwarman@csu.edu.au}
\and
\IEEEauthorblockN{Muhammad Ashad Kabir}
\IEEEauthorblockA{\textit{School of Computing, Mathematics and Engineering} \\
\textit{Charles Sturt University}\\
NSW, Australia \\
akabir@csu.edu.au}
}

\maketitle

\begin{abstract}
Fake news has emerged as a critical global issue, magnified by the COVID-19 pandemic, underscoring the need for effective preventive tools. Leveraging machine learning, including deep learning techniques, offers promise in combatting fake news. This paper goes beyond by establishing BERT as the superior model for fake news detection and demonstrates its utility as a tool to empower the general populace. We have implemented a browser extension, enhanced with explainability features, enabling real-time identification of fake news and delivering easily interpretable explanations. To achieve this, we have employed two publicly available datasets and created seven distinct data configurations to evaluate three prominent machine learning architectures. Our comprehensive experiments affirm BERT's exceptional accuracy in detecting COVID-19-related fake news. Furthermore, we have integrated an explainability component into the BERT model and deployed it as a service through Amazon's cloud API hosting (AWS). We have developed a browser extension that interfaces with the API, allowing users to select and transmit data from web pages, receiving an intelligible classification in return. This paper presents a practical end-to-end solution, highlighting the feasibility of constructing a holistic system for fake news detection, which can significantly benefit society.
\end{abstract}


\section{Introduction}
Fake news is known by many interchangeable names, with the main two being the terms ``Fake news" itself and ``Misinformation"~\cite{zhou2020survey,covid19_survey}. These terms are used to mean false or misleading information shared to deceive an individual, group, or population into believing something that is clearly not true, often with political motivations with the intention of damaging public trust. In the context of this article, we refer to all of this as fake news~\cite{lazer_etal_2018}.

Social networks provide a platform that millions of people around the world use to communicate and share information on a daily basis. However, especially at a time of global crisis such as during the COVID-19 pandemic, the amount of Fake news, being shared is staggering. Global statistics indicate that 74\% people are very concerned about the amount of fake news they have seen during the pandemic~\cite{watson_2020stat1}, and furthermore, studies have shown that more than 50\% of all social media users have spread fake news knowingly or unknowingly~\cite{watson_2020stat2}.

Further research shows that fake news is not more likely to be shared by robots or artificial intelligence, people were found to be more likely to spread fake news~\cite{vosoughi_roy_aral_2018}. A potential reason for sharing fake news is identified as our cognitive biases, more specifically our memory biases, and a phenomenon known as the ``false memory effect"~\cite{britt_rouet_blaum_millis_2019}. Additionally, there has been shown to be a large disconnect between what people believe and what types of fake news they will share, furthering the idea that people share this information irrationally~\cite{pennycook_rand_2021}.

The magnitude of this problem has been underscored by the World Health Organization, which categorizes the proliferation of fake news during the COVID-19 pandemic as an ``Infodemic"~\cite{thelancetinfectiousdiseases_2020}. They explicitly emphasize that an Infodemic can pose as much risk to public health and security as the virus itself.

Given the clarity the World Health Organisation provides on the enormity of this issue, along with the consideration of how easily people with the best intentions can share fake news, the development of a tool that can provide a person with an understanding of what they are reading with regard to its legitimacy and validity is vital. Furthermore, considering how fake news can be used maliciously with the intent to harm public trust and cause unrest, it is important that people have the ability to protect themselves to prevent a continued deterioration of the public's understanding of the pandemic among other topics.

\begin{figure*}[!t]
     \centering
     \begin{subfigure}[b]{0.24\textwidth}
         \centering
         \includegraphics[width=\textwidth]{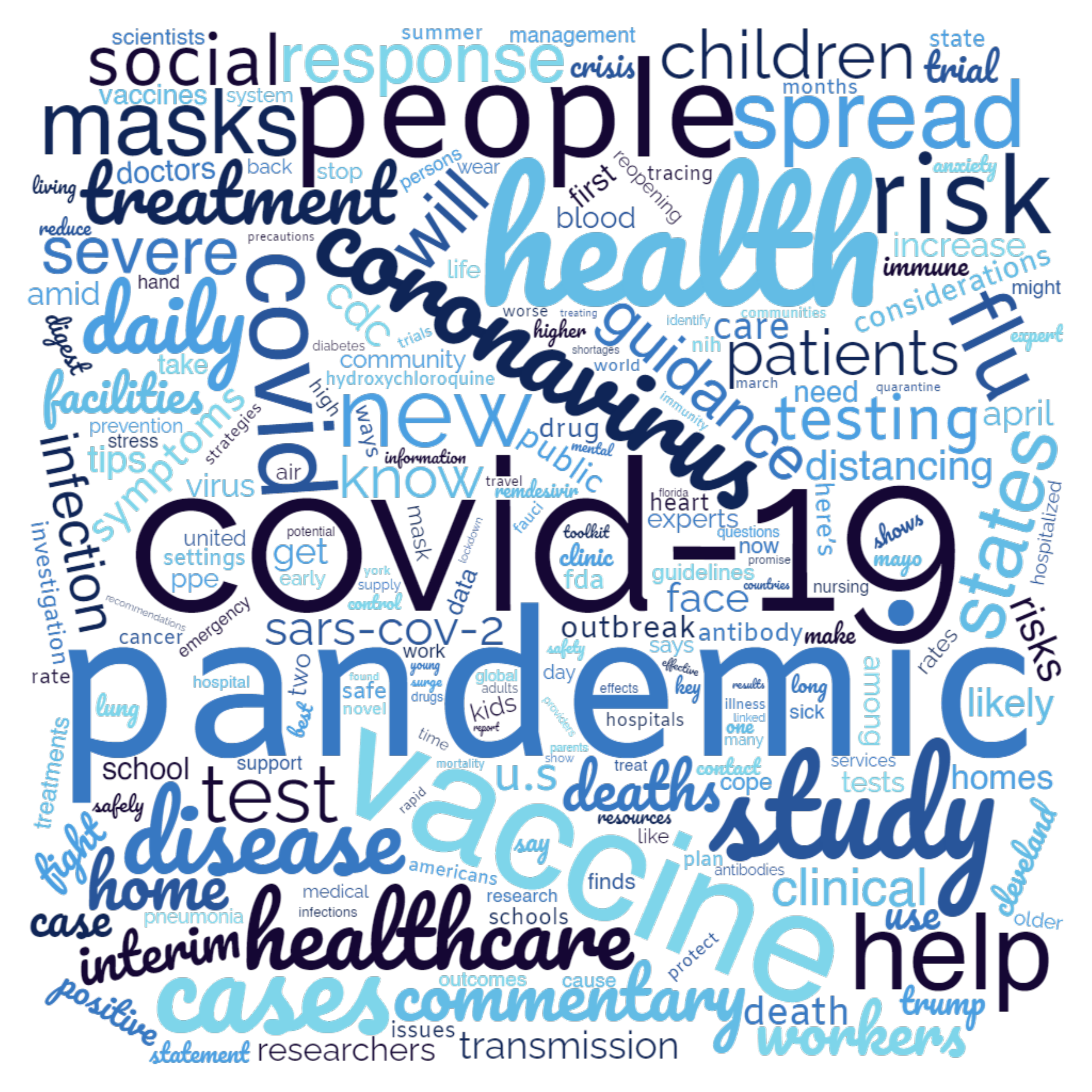}
         \caption{CoAID -- real news}
         \label{fig:CoAID Real News - Word Cloud}
     \end{subfigure}
     \begin{subfigure}[b]{0.24\textwidth}
         \centering
         \includegraphics[width=\textwidth]{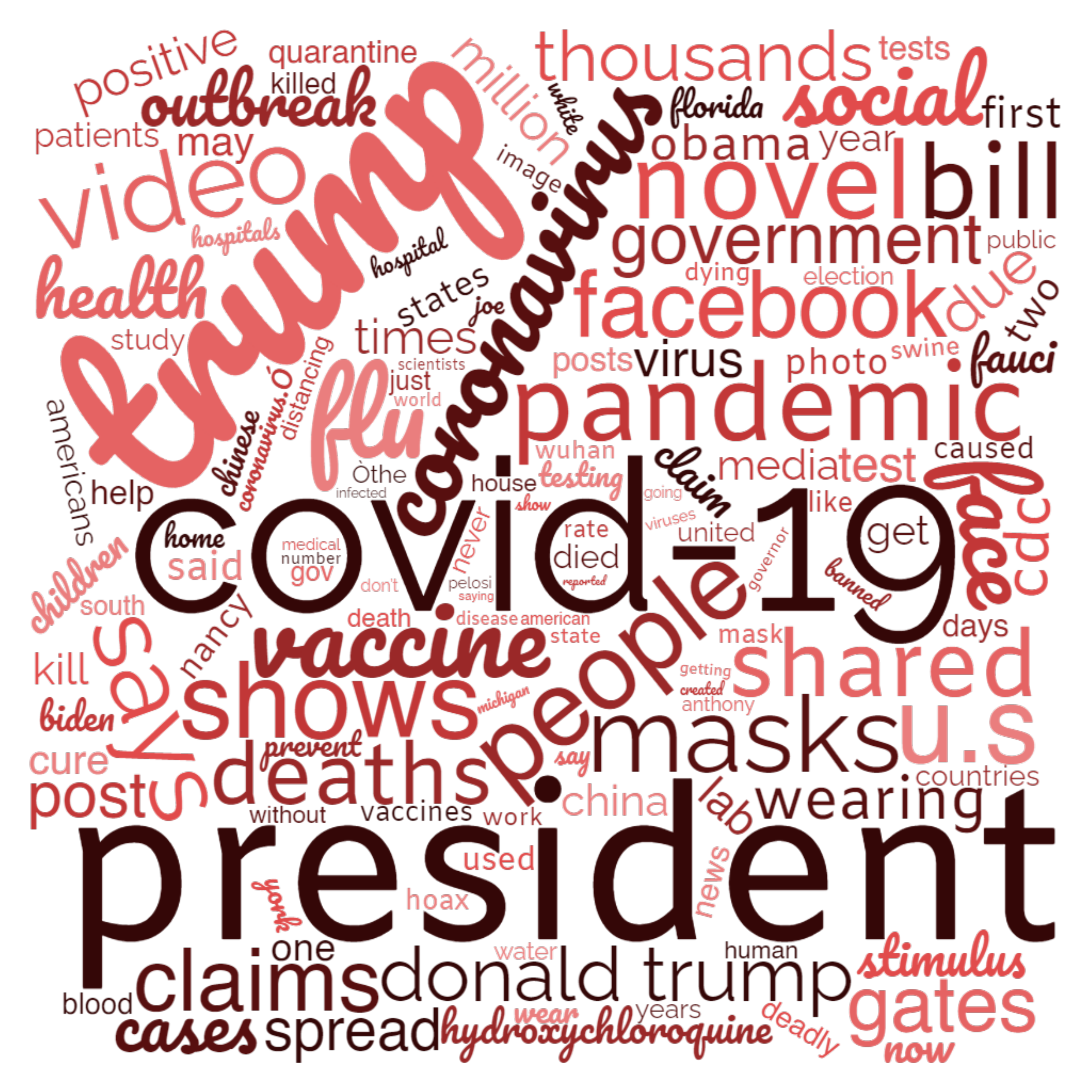}
         \caption{CoAID -- fake news}
         \label{fig:CoAID Fake News - Word Cloud}
     \end{subfigure}
    \begin{subfigure}[b]{0.24\textwidth}
         \centering
         \includegraphics[width=\textwidth]{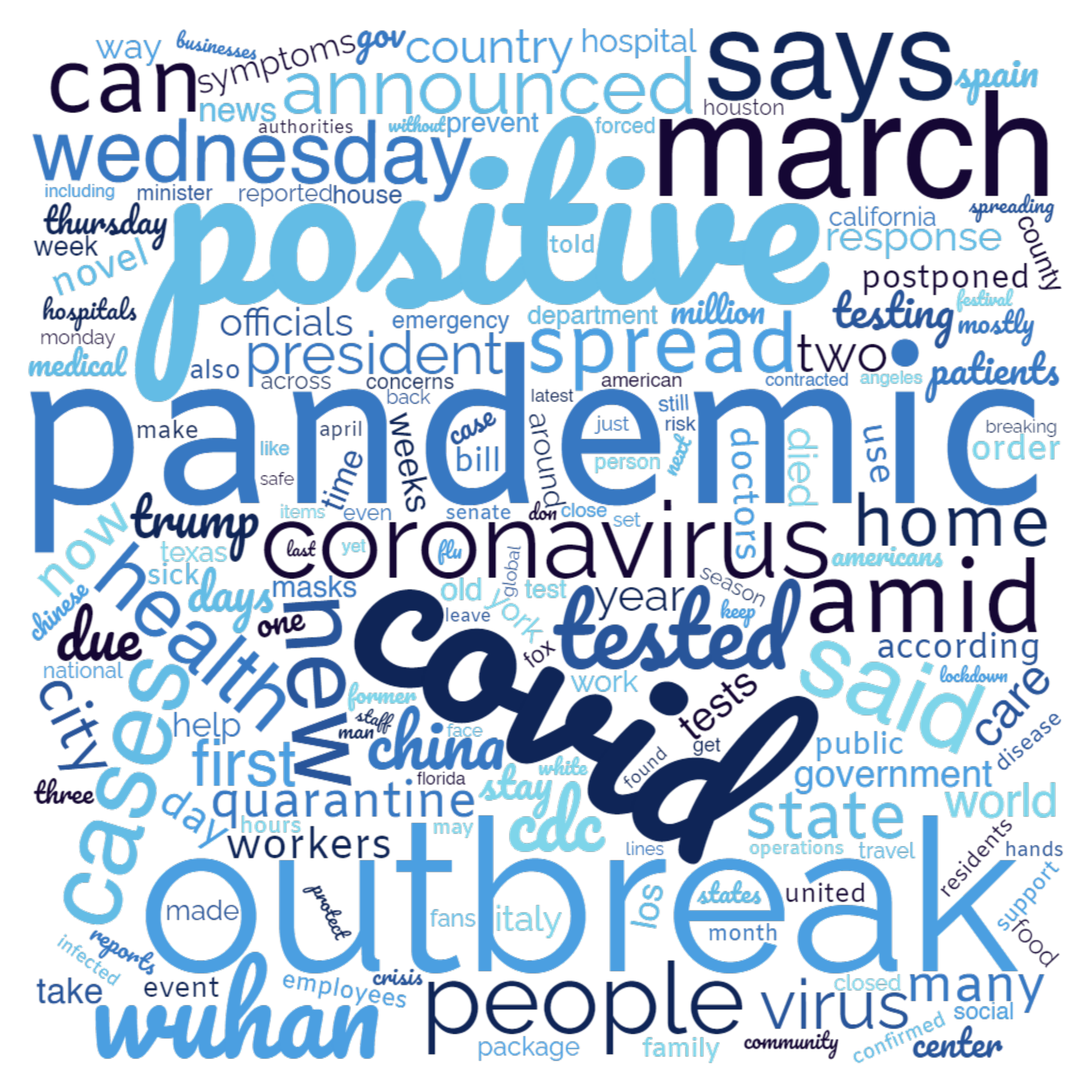}
         \caption{C19-Rumor -- real news}
         \label{fig:Rumour Real News - Word Cloud}
     \end{subfigure}
     \begin{subfigure}[b]{0.24\textwidth}
         \centering
         \includegraphics[width=\textwidth]{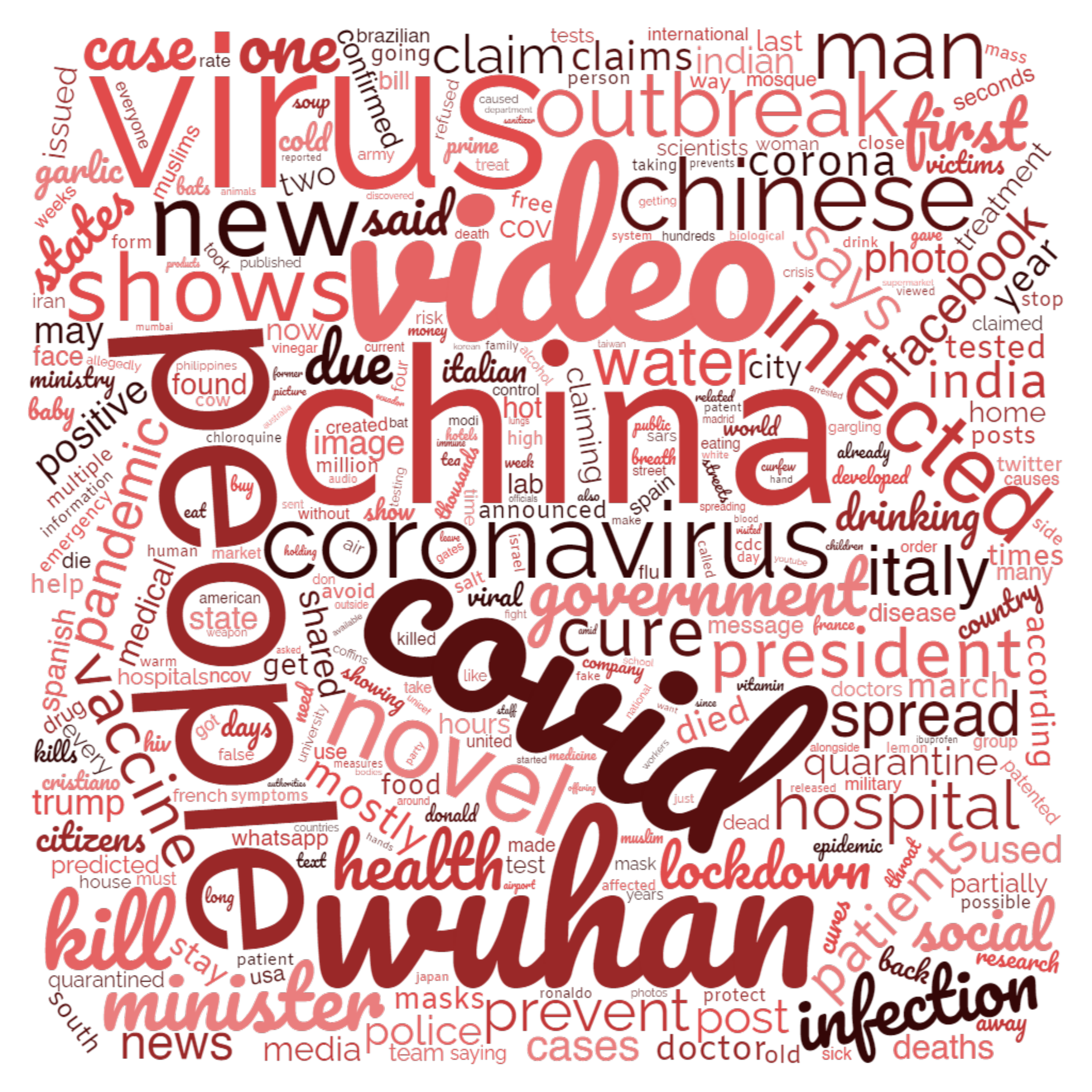}
         \caption{C19-Rumor -- fake news}
         \label{fig:Rumour Fake News - Word Cloud}
     \end{subfigure}
        \caption{Word clouds generated from COVID-19 news datasets}
        \label{fig:datasetWordClouds}
\end{figure*}
Significant research has been conducted in this area recently, with machine learning (ML) approaches being the most widely implemented~\cite{covid19_survey}. Incorporating explainability~\cite{Samek2017} could further allow us to trace the predictions generated by these ML approaches, in particular deep learning (DL) models which are considered as ``black box", and provide a contextual understanding of why a particular classification is made.
Therefore, explainability in fake news detection could significantly increase user trust~\cite{Ayoub2021} and would provide a concrete understanding to the user why an article is fake and could potentially lead to a reduction in the sharing of~fake~news. 


This paper aims to develop an explainability tool/application that is embedded directly into a Google Chrome extension. In particular, this paper makes the following three major contributions:
\begin{itemize}
  \item We have conducted an extensive empirical evaluation of state-of-the-art ML algorithms to train a fake news classification model using two different datasets with seven configurations. 
  
  \item We have identified the most prominent explainability techniques and discussed their suitability for developing a web-based application.
  
  \item We have implemented a web application as a Google Chrome extension using the best-performed ML model and the most suitable explainability technique to demonstrate the suitability and usefulness of our approach. 
\end{itemize}

\section{Related Work}
The current landscape of tools for detecting fake news reveals several limitations and gaps in addressing the critical need for accessible and explainable solutions. CoVerifi~\cite{kolluri_murthy_2021} is a functional application that provides accurate classifications and human generation scores for COVID-19-related news articles, but it lacks genuine explainability techniques, leaving users without a comprehensive understanding of the reasoning behind the classification. However, a more fundamental issue highlighted by dEFEND~\cite{Shu2019} is the scarcity of tools accessible to end-users for detecting fake news.

A notable example of such limitations is FakerFact~\cite{ma_towey_yueh_2021}, a Chrome extension that analyzes and verifies fake news by URL. Although it offers classification percentages in various areas, it falls short of providing direct explainability. Similarly, SEMiNExt~\cite{shams_hoqueSEMINEXT_2021} analyses user search terms for potential fake news content without explaining or classifying actual news articles. While there are tools~\cite{Ayoub2021} like xFake~\cite{yang_pentyala_mohseni_du_yuan_linder_ragan_ji_hu_2019} that demonstrate the potential for sentiment and linguistic analysis with explainable outputs, they often have limitations, like being restricted to specific websites, as seen in xFake's compatibility with PolitiFact~\cite{poynterinstitute_2021} only.

Despite these limitations, tools like Bunyip~\cite{sawant_2020} provide promising visual explainability outputs and classifications for human-generated text, showing that similar applications can be developed. However, these tools do not directly address traditional fake news detection, although they serve as proof of concept for the feasibility of creating user-friendly and explainable applications.

In summary, the existing tools fall short of providing comprehensive and user-friendly solutions for detecting fake news, particularly identifying COVID-19-related fake news with explainability. There is a significant gap between the state-of-the-art studies on machine learning and explainability techniques for fake news, and the ideal end-user tools that offer both accurate classifications and clear interpretable explanations. While the above-discussed tools demonstrate progress, they underscore the necessity for a holistic approach that provides a user-friendly experience coupled with meaningful explainability to empower users in identifying and understanding fake news.

\begin{table*}[!t]
    \centering
    \caption{Dataset Configurations}
    \label{tab:config}
    \begin{tabular}{lllrrrc}
    \hline
        \multicolumn{2}{c}{\multirow{2}{*}{Dataset}} & \multirow{2}{*}{Split} & \multicolumn{2}{c}{Class} & \multirow{2}{*}{Total} & \multirow{2}{*}{Configuration}  \\
        \cline{4-5}
          & & & True (Real) & False (Fake) & \\
          \hline\hline
          \multirow{7}{*}{CoAID} & \multirow{3}{*}{Original} & Training (70\%) & 2426 & 634 & 3060 & \multirow{3}{*}{C1}\\
                                & & Validation (20\%) & 681 & 193 & 874 &\\
                                & & Testing (10\%) & 349 & 89 & 438 &\\
           \cline{2-7}
                                 & \multirow{3}{*}{Augmented} & Training & 2419 & 2419 & 4838 & \multirow{3}{*}{C2}\\
                                & & Validation & 688 & 688 & 1376 &\\
                                & & Testing & 349 & 349 & 698 &\\
          \hline
          \multirow{7}{*}{C19-Rumor} & \multirow{3}{*}{Original} & Training & 452 & 2137 & 2589 & \multirow{3}{*}{C3}\\
                                & & Validation & 145 & 595 & 740 &\\
                                & & Testing & 62 & 308 & 370 &\\
           \cline{2-7}
                                 & \multirow{3}{*}{Augmented} & Training & 2117 & 2117 & 4234 & \multirow{3}{*}{C4}\\
                                & & Validation & 615 & 615 & 1230 &\\
                                & & Testing & 308 & 308 & 616 &\\
         \hline
          \multirow{6}{*}{Cross} & \multirow{3}{*}{\makecell[t l]{CoAID train and validation\\C19-Rumor test}} & Training & 2779 & 2779 & 5558 & \multirow{3}{*}{C5}\\
                                & & Validation & 677 & 677 & 1354 &\\
                                & & Testing & 659 & 659 & 1318 &\\
           \cline{2-7}
                                 & \multirow{3}{*}{\makecell[t l]{C19-Rumor train and validation\\CoAID test}} & Training & 2443 & 2443 & 4886 & \multirow{3}{*}{C6}\\
                            & & Validation & 597 & 597 & 1194 &\\
                            & & Testing & 916 & 916 & 1832 &\\
           \hline
           \multicolumn{2}{l}{\multirow{3}{*}{Merged}} & Training & 2870 & 2779 & 5649 & \multirow{3}{*}{C7}\\
                            & & Validation & 836 & 778 & 1614 &\\
                            & & Testing & 409 & 399 & 808 &\\
         \hline
          
          \hline
    \end{tabular}
\end{table*}
\section{Materials and Methods}
\subsection{Dataset Pre-processing and Configurations}
We have used the CoAID (Covid-19 heAlthcare mIsinformation Dataset) \cite{cui2020coaid} with 5216 total news items and the C19-Rumor (A COVID-19 Rumor Dataset) dataset \cite{cheng2021covid} with a total of 4129 news items. Both datasets have dedicated segments for news headlines and news articles about COVID-19. 
We selected these two datasets given their spread of real and fake news complement each other, with CoAID heavily weighted towards true or real news and C19-Rumor heavily weighted towards false or fake news. This alternate weighting allows us to test the datasets individually, with augmentation, and in combination, to assess the impacts of different class weightings on the data.

Fig.~\ref{fig:datasetWordClouds} illustrates the word cloud of real and fake news for both datasets. It allows for a direct comparison between the two datasets. From these word clouds, we can deduce that while COVID-19 and its variants are the predominant terms used within the CoAID dataset, the key terms within the C19-Rumor dataset are more aligned with words such as pandemic, outbreak, China, and Wuhan. This observation suggests that the CoAID dataset, given its focus on health and medical-related news, is unlikely to contain information related to the outbreak, videos, or Wuhan specifically.


We created seven configurations using the two named datasets for an extensive experimental evaluation.
These configurations, as well as the training, validation, and test splits, are outlined in Table \ref{tab:config}. The objectives of configurations C1 and C3 are to establish baselines for each of the two datasets. Configurations C2 and C4 aim to assess whether data augmentation, particularly considering the limited dataset size, offers any advantages in terms of accuracy or if it adversely affects classification. C5 and C6 are employed to investigate two aspects: firstly, whether the datasets are representative of each other, and secondly, how well a model developed using 2020 COVID-19 fake news performs on news stories from 2021, and vice versa. Finally, the purposes of C7 (merged dataset) are to improve model robustness, increase generalisation, and mitigate the limitations associated with individual datasets such as imbalances in class distribution, lack of coverage for specific topics, or biases in data collection, which can ultimately enhance the accuracy and effectiveness of fake news detection systems.

\subsection{Machine Learning Techniques}
\label{sec:mltech}
In this study, we employed a baseline CNN~\cite{9451544} model, and two advanced models, BERT~\cite{Devlin2019BERTPO} and Bi-LSTM~\cite{kolluri_murthy_2021} which have demonstrated high efficiency in fake news detection according to prior research~\cite{Ayoub2021, kolluri_murthy_2021, colruiz-segura-bedmar_2020}. Both BERT and Bi-LSTM have shown strong performance with small datasets~\cite{khan_khondaker_afroz_uddin_iqbal_2021, EzenCan2020ACO}, which is crucial given the small size of our sourced datasets. In contrast, a CNN model is a more basic algorithm type compared to the first two, providing a valuable baseline to assess what a simpler algorithm can achieve with the same dataset.

BERT~\cite{Devlin2019BERTPO} holds immense promise for fake news detection due to its ability to comprehend the nuances of language and context. By pre-training on a massive corpus of text, BERT becomes adept at understanding the subtle linguistic cues that often distinguish fake news from genuine content. Its bidirectional architecture allows it to capture relationships between words, making it highly effective in discerning the contextual intricacies that fake news articles often employ to deceive readers. Additionally, BERT's fine-tuning capability enables it to adapt to specific datasets, thereby enhancing its accuracy in identifying misleading or fabricated information. As fake news continues to pose a significant challenge, BERT's natural language processing prowess positions it as a valuable tool in the ongoing fight against misinformation and disinformation.

Bi-LSTM~\cite{kolluri_murthy_2021}, on the other hand, is a recurrent neural network, that specializes in capturing sequential dependencies in text. This makes it particularly effective at discerning subtle linguistic patterns within shorter pieces of text and excels at analyzing the structural flow of information within news articles.
\begin{figure*}[!t]
     \centering
     \includegraphics[width=.8\textwidth]{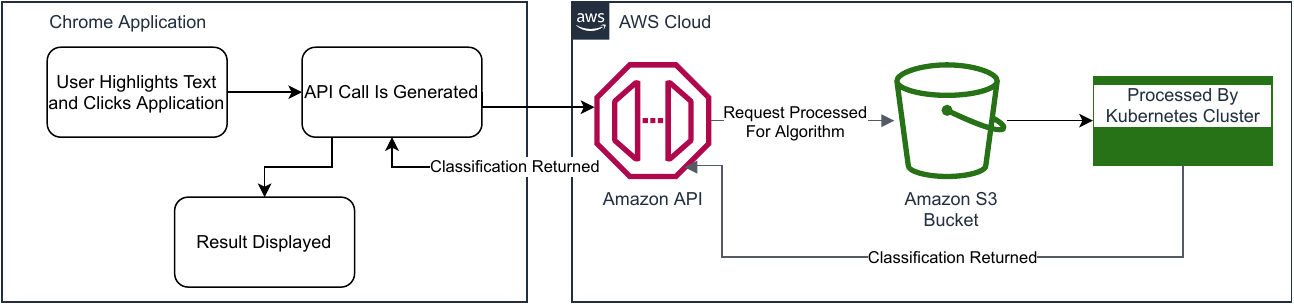}
     \caption{Web application architecture}
     \label{fig:chromarch}
\end{figure*}
\subsection{Explainable Techniques}
SHAP (SHapley Additive exPlanations)~\cite{SHAPpaper} and LIME (Local Interpretable Model-Agnostic Explanations)~\cite{LIMEPaperRibeiro2016} are two popular explainable machine learning techniques used in the context of fake news detection to provide insights into model predictions and make the decision-making process more transparent. In this study, we employed SHAP as it holds several advantages over LIME when it comes to explaining the predictions of machine learning models. One key advantage is the global interpretability that SHAP offers. Unlike LIME, which provides local explanations for individual predictions, SHAP calculates feature importance consistently across all possible feature combinations~\cite{SHAPpaper}. This means that SHAP gives a holistic view of how each feature impacts model predictions across the entire dataset, allowing for a more comprehensive understanding of the model's behavior. Additionally, SHAP is grounded in cooperative game theory, providing a mathematically rigorous framework for explaining the contributions of each feature. This makes SHAP particularly useful when it needs to identify overarching patterns and relationships within data, which is often crucial in applications like fake news detection, where understanding global linguistic and contextual patterns is essential for model transparency and improvement.

\subsection{Web Application Architecture}
Our web application's architecture, depicted in Fig.~\ref{fig:chromarch}, comprises two main components: a server component hosted on the AWS (Amazon Web Services) cloud and a client component implemented as a Chrome plugin.

The server for our application is hosted on an S3 (Simple Storage Service) Bucket within AWS\footnote{\url{https://aws.amazon.com/}}. This bucket serves as a storage location, providing access to the files required to run the algorithm. The server hosts a trained machine learning model (Section \ref{sec:mltech}) and the request handler for the application. To publish this on AWS and generate the necessary components for the API, we used Cortex\footnote{\url{https://github.com/cortexlabs/cortex}}. Cortex is a free application that employs Docker images to automatically create Kubernetes clusters on AWS. A Kubernetes cluster consists of a set of functions from AWS that facilitate communication among the nodes, each representing a specific feature or function. This approach allows us to generate an API and a public access point for the application without the need to manually configure each feature individually.
Once the API is accessible, it is connected using JavaScript within the client component (i.e., Chrome extension). This involves sending a request to the previously uploaded request handler within the S3 bucket, which generates a prediction and a set of force values representing explainability. These results are then returned to the Chrome extension.

The client component (Chrome plugin) is built using HTML, with JavaScript providing the essential functionality for transmitting and processing text. JavaScript processes a request by reading the user-selected data (highlighted text) from the active webpage when the user clicks the application icon. Subsequently, the extension converts the response (classification result with an explanation) from the server into a readable format, applies CSS formatting to provide color-coding for the explainability element, and presents the results to the user.

\section{Results and Discussion}
\subsection{ML Models Evaluation}
\begin{figure*}
     \centering
     \begin{subfigure}[b]{0.24\textwidth}
         \centering
         \includegraphics[width=\textwidth]{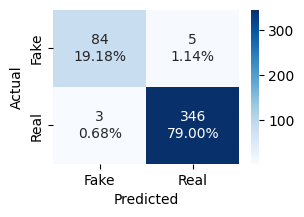}
         \caption{Configuration 1}
         \label{fig:C1}
     \end{subfigure}
     \hfill
     \begin{subfigure}[b]{0.24\textwidth}
         \centering
         \includegraphics[width=\textwidth]{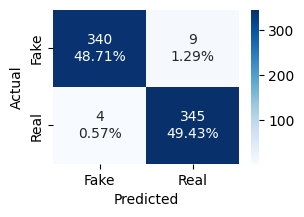}
         \caption{Configuration 2}
         \label{fig:C2}
     \end{subfigure}
     \hfill
     \begin{subfigure}[b]{0.24\textwidth}
         \centering
         \includegraphics[width=\textwidth]{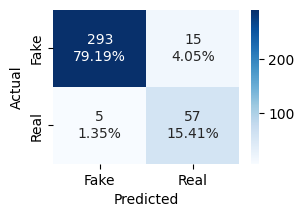}
         \caption{Configuration 3}
         \label{fig:C3}
     \end{subfigure}
     \hfill
     \begin{subfigure}[b]{0.24\textwidth}
         \centering
         \includegraphics[width=\textwidth]{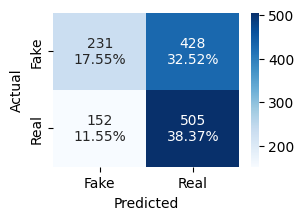}
         \caption{Configuration 4}
         \label{fig:C4}
     \end{subfigure}
     
     \begin{subfigure}[b]{0.25\textwidth}
         \centering
         \includegraphics[width=0.90\textwidth]{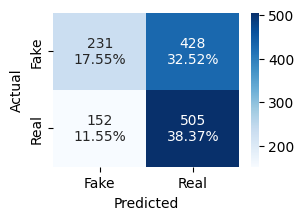}
         \caption{Configuration 5}
         \label{fig:C5}
     \end{subfigure}
     \hfill
     \begin{subfigure}[b]{0.25\textwidth}
         \centering
         \includegraphics[width=\textwidth]{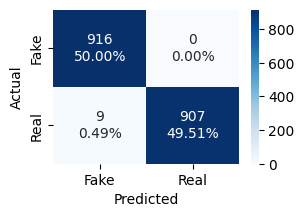}
         \caption{Configuration 6}
         \label{fig:C6}
     \end{subfigure}
     \hfill
     \begin{subfigure}[b]{0.25\textwidth}
         \centering
         \includegraphics[width=\textwidth]{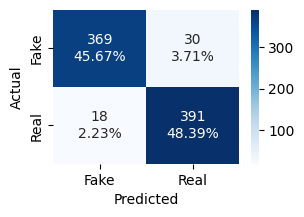}
         \caption{Configuration 7}
         \label{fig:C7}
     \end{subfigure}
        \caption{Confusion matrix of the best model of each configuration}
        \label{fig:confusionmatrixs}
\end{figure*}

The results of running the three models (BERT, Bi-LSTM, and CNN) for seven configurations (C1 to C7) are presented in Table \ref{tab:modelResults}.
\begin{table}[!t]
    \centering
    \caption{Classification Results}
    \label{tab:modelResults}
    \begin{tabular}{llcccc}
    \hline
    Config.                   & Model name & Precision & Recall & F1             & Accuracy \\ \hline \hline
    \multirow{3}{*}{C1} & BERT       & 98.16     & 98.17  & 98.17          & \textbf{98.17}     \\ 
                              & Bi-LSTM        & 92.77     & 92.92  & 92.81         & 92.92   \\ 
                              & CNN       & 93.72     & 93.84  & 93.74          & 93.84    \\ 
                              \hline
    \multirow{3}{*}{C2} & BERT & 98.15 & 98.14 & 98.14 & \textbf{98.14} \\ 
                              & Bi-LSTM        & 85.85     & 81.09  & 80.47          & 81.09    \\  
                              & CNN       & 86.55     & 83.24  & 82.85          & 83.24    \\ 
                              \hline
    \multirow{3}{*}{C3} & BERT & 95.11 & 94.59 & 94.75 & \textbf{94.59} \\ 
                              & Bi-LSTM     & 87.42     & 88.38  & 87.14 & 88.38    \\  
                              & CNN       & 88.30     & 88.38  & 88.34          & 88.38    \\ 
                              \hline
    \multirow{3}{*}{C4} & BERT & 76.19 & 75.97 & 75.92 & \textbf{75.97} \\ 
                              & Bi-LSTM     & 75.41     & 72.73  & 71.99 & 71.99    \\  
                              & CNN       & 73.14     & 61.80  & 56.47      & 61.80    \\ 
                              \hline
    \multirow{3}{*}{C5} & BERT & 56.23 & 54.78 & 51.98 & 54.78 \\ 
                              & Bi-LSTM     & 51.28     & 50.91  & 47.12 & 50.91    \\  
                              & CNN       & 57.06     & 55.84  & 53.85   & \textbf{55.84}    \\ 
                              \hline
    \multirow{3}{*}{C6} & BERT & 99.41 & 99.40 & 99.40 & \textbf{99.40} \\ 
                              & Bi-LSTM   & 50.00 & 50.00 & 42.44 & 50.00 \\  
                              & CNN       & 46.95 & 49.13 & 38.07 & 49.13 \\ 
                              \hline
    \multirow{3}{*}{C7} & BERT & 94.10 & 94.06 & 94.06 & \textbf{94.06} \\ 
                              & Bi-LSTM     & 86.76     & 86.76  & 86.76 & 86.76    \\  
                              & CNN        & 88.07     & 88.00  & 87.98          & 88.00    \\ 
                              \hline
                              
                              \hline
    \end{tabular}
\end{table}
For configuration C1, all three models consistently perform well, achieving the highest average and displaying the most consistent scores among all the tests. However, it is important to consider whether this high accuracy might be influenced by the imbalanced nature of the dataset. This consideration is quickly addressed by examining Fig.~\ref{fig:C1}, which displays the BERT confusion matrix. Here, only five fake labels were misclassified compared to the three from the real set. This suggests that the dataset's imbalanced nature is unlikely to have significantly influenced the accuracy. Instead, it appears that the dataset contains well-separated data that the models were able to effectively distinguish.

For configuration C2, it is worth noting that while BERT maintains a consistent score, with only a minor drop of 0.03\%, both the Bi-LSTM and CNN models experience significant decreases in accuracy, exceeding 10\%. This drop in accuracy suggests that oversampling is ineffective for this dataset, and it indicates that BERT is more resilient to the challenges posed by oversampled data. Fig.~\ref{fig:C2} displays the relationship between predicted labels and true labels, which closely resembles that of C1. This further confirms that the accuracy achieved in C1 was not solely due to the imbalanced dataset. Despite the slight reduction in accuracy, this configuration strengthens the case that BERT is the optimal model for this particular scenario.

For configuration C3, we observe lower overall accuracy compared to C1. This suggests that the examples in this dataset are more challenging to distinguish than those in the first dataset. More challenging data can be both a positive and negative aspect. On the one hand, it demonstrates that the algorithm can still differentiate between the examples to a significant extent. On the other hand, the lower accuracy indicates that the algorithms face greater difficulty in establishing connections between textual elements. Additionally, when we examine Fig.~\ref{fig:C3}, we notice that the optimal algorithm's accuracy varies mainly when classifying real data. This variance can be attributed to the overall scarcity of real news in this dataset, resulting in a 20.83\% misclassification rate for real news.

Interestingly, the trend observed in C2 continues in C4, with a significant decrease in accuracy across all three models. BERT experiences the most substantial accuracy drop when compared to C3, which can be partly attributed to its higher initial accuracy. Across all three models, the average drop in accuracy is 20.53\%, a sharp increase compared to the average drop of 7.49\% in C2. This significant increase in accuracy loss suggests that this second dataset is less robust than the first. When comparing Fig.~\ref{fig:C4} to Figure \ref{fig:C2}, the general trend in accuracy is evident. However, Fig.~\ref{fig:C4} displays notably lower accuracy when dealing with what was initially the minority class. Despite the considerable accuracy drop, BERT outperforms the other models once again, strengthening the argument that it is the overall best model.

Examining the cross-dataset evaluation results depicted in Fig.~~\ref{fig:C5} and Fig.~\ref{fig:C6}, we observe an interesting aspect in the results of C5 and C6. Models trained on the CoAID dataset performed poorly on the C19-Rumor dataset (Fig.~\ref{fig:C5}). Conversely, the best model trained on the C19-Rumor dataset excelled when tested on the CoAID dataset (Fig.~\ref{fig:C6}). This phenomenon suggests that the C19-Rumor dataset effectively represents the characteristics of the CoAID dataset, but the reverse is not necessarily true. This relationship is also partly reflected in the word cloud analysis presented in Fig.~\ref{fig:datasetWordClouds}.

In C7, we consistently observe a high level of accuracy, reaffirming the primary objective of this configuration. This objective, centered around testing the models' ability to handle an expanded dataset covering various subcategories, has been convincingly validated. Once again, BERT outperforms the other models in terms of accuracy, as evident in Table~\ref{tab:modelResults}. Fig.~\ref{fig:C7} illustrates a minimal number of misclassified inputs.
The significant advantage of witnessing C7 perform exceptionally well lies in its confirmation that any marginal reduction in accuracy can be attributed to the models' enhanced comprehension of the broader subject of COVID-19. This heightened understanding is a direct result of the merged dataset, which thoughtfully addresses issues like class distribution imbalances, topic coverage limitations, and data collection biases present in individual datasets. As a result, the merged dataset naturally enriches the pool of COVID-19-related data, ultimately fostering the accuracy and effectiveness of fake news detection systems.

\begin{figure*}[!t]
     \centering
     \begin{subfigure}[b]{\textwidth}
         \centering
         \includegraphics[width=\textwidth]{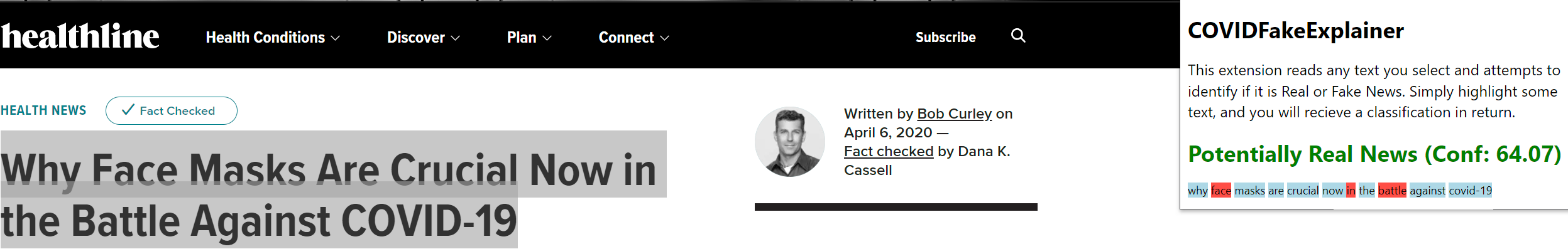}
         \caption{Real news}
         \label{fig:RealNewsExampleFull}
     \end{subfigure}
     \begin{subfigure}[b]{\textwidth}
         \centering
         \includegraphics[width=\textwidth]{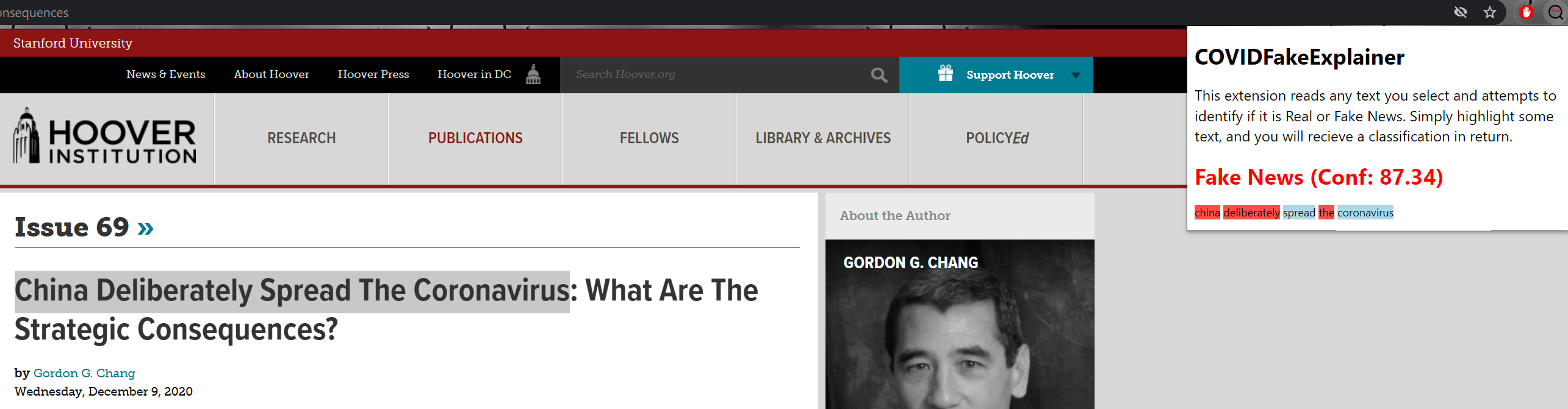}
         \caption{Fake news}
         \label{fig:FakeNewsExampleFull}
     \end{subfigure}
        \caption{Chrome extension is providing classification results with explanation for a selected text in webpage}
        \label{fig:fullappscreenshots}
\end{figure*}
\subsection{Web Application Evaluation}
Fig.~\ref{fig:fullappscreenshots} demonstrates a real-time use of our web application, encompassing the article content and highlighting the specific line within that article that is undergoing classification. Notably, there is currently no similar application except for CoVerifi~\cite{kolluri_murthy_2021}, which necessitates copying the text and leaving the active page or site to generate a response. In contrast, our tool offers a straightforward highlight-and-click function, eliminating the need for copying or additional buttons to produce results, as required by tools that redirect to external pages. Furthermore, our tool grants users the flexibility to select any text they desire, providing complete control over the application's inputs.

\section{Conclusion}
This paper presents a comprehensive pipeline encompassing the entire process, from training machine learning models to prototype implementation, for the detection of fake news with explainability. Our study demonstrates the potential of combining machine learning and explainability techniques to create a web application tailored for detecting COVID-19-related fake news. Through an extensive empirical analysis, we evaluated the performance of three prominent machine learning algorithms for text classification across seven distinct configurations, employing two distinct datasets. The results indicate that BERT emerges as the optimal choice for COVID-19 fake news classification. Moreover, we critically examined the two leading explainability visualization techniques, offering insights into their respective advantages and limitations. Finally, we developed a prototype web application in the form of a Chrome extension. 
This approach is highly adaptable and can be extended beyond COVID-19 fake news detection, for instance, to classify text or misinformation in a variety of domains. The only requirement for this adaptation is the replacement of the model that serves as the foundation for the application.

In the future, we plan to further improve the application's robustness by incorporating a larger and more comprehensive dataset. Additionally, we intend to conduct thorough evaluations to assess the usability and performance of the application.

\bibliographystyle{IEEEtran} 
\bibliography{references}

\end{document}